\documentclass[10pt,journal,cspaper,compsoc]{IEEEtran}

\usepackage{ifpdf}

 \ifpdf

 \else

 \fi

\ifCLASSOPTIONcompsoc
\else
\fi

\ifCLASSINFOpdf
   \usepackage[pdftex]{graphicx}
\else
\fi

\usepackage[cmex10]{amsmath}

\usepackage{algorithmic}

\usepackage{array}

\usepackage{mdwmath}
\usepackage{mdwtab}

\usepackage{eqparbox}

\usepackage[tight,normalsize,sf,SF]{subfigure}

\usepackage{amssymb}

\newtheorem{theorem}{Theorem}[section]

\newtheorem{definition}{Definition}[section]

\newtheorem{remark}{Remark}[section]

\begin{document}

\title{Some considerations about Java implementation of two provably secure pseudorandom bit generators}

\author{Antonio~Corbo~Esposito~ and~ Fabiola~Didone
\IEEEcompsocitemizethanks{\IEEEcompsocthanksitem A. Corbo Esposito and F. Didone are with the Universit\'a 
degli studi di Cassino e del Lazio Meridionale,  Italia.\protect\\}

\thanks{}}

\IEEEcompsoctitleabstractindextext{%
\begin{abstract}
\boldmath
The quest for a cryptographically secure pseudorandom bit generator (PRBG) was initiated long ago \cite{Blum}  \cite{BlumMicali}, and for a long time the proposed pseudorandom generators were very slow \cite{Sidorenko}.
(INSERIRLO IN BIBLIOGRAFIA). More recently some "provably secure" PRBG capable to achieve a throughput rate greater than 1Mbit/sec has been proposed in \cite{RSAPRG} \cite{QUAD1}. We noticed, anyway, the absence of Java implementations of such PRBGs, provably due to poor expected values for throughput rate.

 In the present paer we show that it is quite easy to write down Java implementations for them, achieving a throughput rae into range $0,5\div 7$ Mbit/sec on very common mobile low-end devices. If moreover a modification we proposed in \cite{Articolo1} is applied, the throughput rate is in the range $5\div 80$ Mbit/sec and can be clearly sufficient for many applications.
\end{abstract}

}

\maketitle

\IEEEdisplaynotcompsoctitleabstractindextext

\IEEEpeerreviewmaketitle

\section{Introduction}

\IEEEPARstart{T}{he} main purpose of this paper is to show that it is easy to realize a software implementation in Java of a CSPRBG (cryptographically secure pseudo random bit generator, see definition \ref{CSPRBG}) and to give a quite precise estimate of the throughput rate achieved.
Since many CSPRBGs are based on the iteration of a one way function we developed some Java code to test the speed of the computation of this function for two well known PRBGs in the literature, namely RSAPRG \cite{RSAPRG} and QUAD \cite{QUAD1},\cite{QUAD2}. In the process of optimization of our code we obtained a clear evidence about the following assertions:
\begin{itemize}
\item currently used algorithms for modular reduction with respect to a fixed modulus (i.e. the computation of the remainder of a division) can be improved up to 20$\%$ by simple algebraic methods (when the bit length $n$ of the modulus is $n\geq 6000$);
\item tha Java BigInteger class has a completely different behaviour under MAC or Windows o.s. with respect to Android o.s. . This almost surely corresponds to a different Java code in the two cases;
\item these Java implementations, without any use of parallel computations achieve a throughput rate in the range 0,5$\div$7 Mbit/sec for PRBG that are "provably secure" on very common low-end devices (PCs, tablets or smartphones). 
\end{itemize}

In cryptographic applications, given that Alice and Bob share a private key, the production of a bits stream by a PRBG  based on this key is completely equivalent to an encryption/decryption communication algorithm. They are equivalent in the sense that the communication algorithm can achieve the same throughput rate of the PRBG and that is as secure as PRBG is. Moreover the communication algorithm has zero latency, i.e. the encryption/decryption time is negligible (see paper \cite{Articolo1}).
The security of a PRBG is however very often established in an empirical way and very rarely we deal with a "provable security".
So we face three main questions:
\begin{enumerate}
\item what is "provable security"?
\item  there exist PRBGs that are provably secure?
\item how efficient they can be?
\end{enumerate}
As for the answer to questions 1) and 2) we have found, in literature, two PRBGs that qualify themselves as "provably secure":
\begin{itemize}
\item RSAPRG described in the paper \cite{RSAPRG};
\item QUAD described in papers \cite{QUAD1} and \cite{QUAD2}.
\end{itemize}
In both cases the provable security consists in some reduction theorem of this type: if the sequence produced is distinguishable (according to some parameters) from a real random sequence, then some well known problem could be solved in less then a certain amount of time (depending on the parameters) and this will be less of the time used by the best known algorithm for this problem.

About the efficiency of these PRBGs, the throughput rates they claim (see section 5 of \cite{RSAPRG}, section 4.5 of \cite{QUAD2} and paper \cite{1}) well fit our results and in some cases we outperform their claims.

In paper \cite{Articolo1}, however, we show how it is possible to modify a CSPRBG so that we can increase the throughput rate up to more than $10\times$, while retaining the essential features of provable security. When we apply this modification to RSAPRG and QUAD we obtain a throughput rate in the range $5\div 80$ Mbit/sec on the same devices considered before. It seems to us that to achieve a throughput in such range for a PRBG based on a "provable secure" basic iteration (moreover in a software implementation not using any parallel computation) is a promising result.

\section{CSPRBGs and "provable security"}

\subsection{Some notations}

In the paper in general it will be clear when some letter is used to denote an integer variable, real variable or some other variable type. Moreover we will denote by:
\begin{itemize}
\item $M(n)$ the (minimum) time to multiply two $n$-bits integers;
\item $M(m,n)$ the time to multiply an $m$-bits integer by an $n$-bits integer;
\item $SQ(n)$ the time to compute the square of  an $n$-bits integer;
\item $mod(N)$ the remainder of a division by $N$;
\item $MR(m,n)$ the time to compute the modular reduction of an $m$-bits integer modulo an $n$-bits integer;
\item $\lfloor x \rfloor$ the floor function, $x\in \mathbb{R}$.
\end{itemize}

Many CSPRBGs have a similar structure. We call them "typical" in the sense of the following definition.
\begin{definition}\label{CSPRBG}
A typical CSPRBG (TCSPRBG for short) is given by:
\begin{enumerate}
\item a (private) seed $x_0\in X$ (some numeric space).
\item (public) data:
\begin{itemize}
\item a one way (usually invertible) function $f:X\to X$;
\item a one way function $g:X\to S$ where $S$ is the set of strings of bits.
\end{itemize}
\end{enumerate}
The basic iteration of the TCSPRBG is the computation:
$$x_{h+1}=f(x_h),\ h\in \mathbb{N}$$
$$y_{h+1}=g(x_{h}),\ h\in \mathbb{N}$$
The output of TCSPRBG is the string $S=y_1||y_2||...$ where $||$ denotes the concatenation.
\end{definition}

\subsection{RSAPRG}
The RSAPRG has been proposed by Steinfeld, Pieprzyk and Wang in \cite{RSAPRG} and is a Micali-Schnorr type PRBG.

Given a RSA modulus $N$ of bitlength $n$ (i.e. $N=p\cdot q$ where $p$ and $q$ are two prime numbers of bitlength $\frac{n}{2}$) and an exponent $e$ such that $GCD(e,(p-1)(q-1))=1$, the one way function $f$ is given by $f(x)=x^emod(N)$ while the function $g$ is given by $g(x)=f(x)\ mod(2^r)$ plus the conversion to a bit string, so that at each iteration the $r$ least significant bits of $x_{h+1}$  are outputted. The seed $x_0\in \mathbb{Z}_{N}$. The maximum number of bits outputted will be fixed to $l$. They call such PRBG a $(n,e,r,l)$-RSAPRG.

The security result for RSAPRG is given by corollary 4.1 of \cite{RSAPRG}) that we report here (see \cite{RSAPRG} for related definitions and details).
\begin{theorem}
{\it For all $n\geq 2^9$, any $(T,\delta)$ distinguisher $D$ for $(n,e,r,l)$-RSAPRG can be converted into a $(T_{INV},\epsilon{INV})$ inversion algorithm $A$ for the $(n,e,r,w)$-CopRSA problem (with $w=3\log(2l/\delta)+5)$ with $T_{INV}=64\cdot(l/\delta)^2n\log(n)\cdot (T+O(l/r\log (e)n^2))$ and $\epsilon_{INV}=\delta/9-4/2^{n/2}$.}
\end{theorem}
Loosely speaking the $(n,e,r,w)$-CopRSA problem is the following one: we are given $x^emod(N)$, $n/2+w$ most significant bits of $x$, $r$ last significant bits of $x$ and we are requested to find $x$. In section 5 of \cite{RSAPRG} it is concretely estimated the minimum time needed to solve $(n,e,r,w)$-CopRSA problem. This is done regarding it as a particular SSRSA (small solution RSA) problem and assuming that the best way to solve it will be either to factorize $N$ or to run a celebrated Coppersmith attack \cite{COPPERSMITH}. In the table 1 of \cite{RSAPRG} authors give some concrete values. For example taking $e=9$ and $n=6144$, at each multiplication $2196$ bits can be sent to output. In this case the RSAPRG can be proved to be safe if moreover the total number of bits outputted is less than $2^{32}$($\sim 500$ Mbytes).

\subsection{QUAD}
This PRBG has been proposed by Berbain, Gilbert and Patarin in \cite{QUAD1}. In this case the seed ${ \bf x}_0=({ x}_{0,1},...,{ x}_{0,n})\in \{0,1\}^n=(\mathbb{Z}_{2})^n$. The (public) data are given by $kn$ multivariate quadratic polynomials $P_1({\bf x}),...,P_{kn}({ \bf x})$ in GF(2) (i.e. ${\bf x}=(x_1,..,x_n)\in (\mathbb{Z}_2)^n$ and the coefficients of polynomials are in $\mathbb{Z}_2$).
The basic iteration is given by:
$${\bf x}_{h+1}=({ x}_{{h+1},1},...,{ x}_{h+1,n})=(P_1({\bf x}_h),...,P_n({\bf x}_h))$$
while:
$${\bf y}_{h+1}=({ y}_{h+1,1},...,{ y}_{h+1,(k-1)n})=(P_{n+1}({\bf x}_h),...,P_{kn}({\bf x}_h))$$
is converted to a string and sent to output.
The security result for QUAD is stated in theorem 4 of \cite{QUAD1} that we report here:
\begin{theorem}
{\it Let $L=\lambda(k-1)n$ the number of keystream bits produced by in time $\lambda T_S$ using $\lambda$ iterations of QUAD construction. Suppose there exists an algorithm $A$ that distinguishes the $L$-bit keystream sequence associated with a known randomly chosen system $S$ and an unknown randomly chosen initial internal state $x\in \{0,1\}^n$ from a random $L$-bit sequence in time $T$ with advantage $\epsilon$. Then there exists an algorithm $C$, which given the image $S(x)$ of a randomly chosen (unknown) $n$-bit value $x$ a randomly chosen $n$-bit to $m$-bit quadratic system $S$ produces a preimage of $S(x)$ with probability at least $\epsilon'=\frac{\epsilon}{2^3\lambda}$ over all possible values of $x$ and $S$ in time upper bounded by $T'$.
$$T'=\frac{2^7n^2\lambda^2}{\epsilon^2}\bigg(T+(\lambda+2)T_S+\log\bigg(\frac{2^7n\lambda^2}{\epsilon^2}\bigg)+2\bigg)+\frac{2^7n\lambda^2}{\epsilon^2}T_S$$}
\end{theorem}

The security result is linked to hardness of MQ problem, i.e. to find a solution to a random choosen multivariate quadratic system. This problem is proved to be NP hard, even for the case of quadratic polynomials over the field GF(2). Furthermore, in contrast to number theoretic problems like integer factorization and discrete logarithms, it is believed to resist quantum computer attacks \cite{2}.
In paper \cite{QUAD1} authors refer to an estimate of Magali Bardet \cite{BARDET} for the best time to solve a generic MQ problem of $kn$ equation in $n$ unknowns using Groebner basis, given by:
$$T(k,n)\simeq \bigg( \bigg( \begin{matrix} {n+1} \\  D \end{matrix}\bigg) \bigg)^{2,37}$$
where:
$$\frac{D}{n}=-k+\frac{1}{2}+\frac{1}{2}\sqrt{2k^2-10k-1+2(k+2)\sqrt{k(k+2)}}$$
For a better understanding we easily obtained the following expansion for $D$ for larger values of $k$:
$$\frac{D}{n}\sim \frac{1}{8k}-\frac{1}{16k^2}+\frac{7}{128k^3}+o\bigg( \frac{1}{k^3}\bigg)$$

To obtain a contradiction time $T'$ has to be smaller than $T(k,n)$ by a factor $\frac{1}{\epsilon'}$, assuming $k=2$, if one wants output a maximum of $2^{40}$ bits a contraddiction is obtained if $n\geq 350$. The authors, however, judging their proofs not optimal, suggest to take for practical applications $k=2$, $n=160$, $L\leq 2^{40}$.

\section{How to implement the basic iteration }

\subsection{Implementation of the basic iteration of RSAPRG}

In this case the basic iteration is given by the computation of $x^emod(N)$, where  $x$ and $N$ have the same bitlength $n$ and $e$ is a low exponent. Following \cite{RSAPRG} pag 9 we assume $n=6144$ and $e=9$ we can compute $x^9mod(N)$ as follows:

\begin{array}{l}
y\leftarrow x\\
\mbox{for $j=1$ to 3}\\
\{x\leftarrow x^2\\
x\leftarrow xmod(N)\}\\
x\leftarrow yx\\
x\leftarrow xmod(N)
\end{array}

The computation requires 3 squares $SQ(n)$, 1 multiplication $M(n)$ and 4 modular reductions $MR(2n,n)$\footnote{An alternative procedure could be exploited by using an algorithm \cite{Bodrato} that directly computes $x^3$ without computing $x^2$. For our purposes (calculation of $x^9$) this procedure anyway should turn out to be (slightly) slower.}.

In the cryptographic range $(10^3\leq n\leq 10^5)$ the most efficient way to perform $M(n)$ is given by a subquadratic algorithm (i.e Karatsuba \cite{KARATSUBA} or Toom Cook \cite{TOOMCOOK}). For reference about these methods we refer to \cite{MCA}. In our case $n=6144$ and we can assume that Karatsuba or Toom Cook are roughly equivalent (the actual behaviour can be different depending on both software and hardware).\\
Since for Karatsuba algorithm:
\begin{equation}
M(n)\simeq3M\bigg(\frac{n}{2}\bigg)
\end{equation}
while for Toom Cook $3$:
\begin{equation}
M(n)\simeq5M\bigg(\frac{n}{3}\bigg)
\end{equation}
this amounts to assume:
\begin{equation}\label{eq1}
M(n)\simeq 3M\bigg(\frac{n}{2}\bigg)\simeq 5M\bigg(\frac{n}{3}\bigg)
\end{equation}
in our application range.\\
Moreover $SQ(n)$ can be roughly estimated by:
\begin{equation}
SQ(n)\simeq \frac{2}{3}M(n)
\end{equation}
(see subsection 1.3.6 of \cite{Zimmermann}).\\
As for modular reduction in our case we have to compute (only) the remainder of a division of an integer of bit length $2n$ division by the fixed modulus $N$ of bit length $n$.

Several algorithms are known to perform this computation, the most famous being the Barrett algorithm and the Montgomery algorithm. In Barrett algorithm the following computations are performed to yield $x\ mod(N)$:

\begin{array}{l}
\hat{Q}\leftarrow \left\lfloor\frac{\left\lfloor\frac{x}{2^n}\right\rfloor \left\lfloor\frac{2^{2n}}{N}\right\rfloor}{2^n} \right\rfloor\\
x\leftarrow x-\hat{Q}N\\
while\ x\geq N\\
\{ x\leftarrow x-N\}
\end{array}

(if $Q=\left \lfloor \frac{x}{N}\right \rfloor$ it can be proved that $Q-2\leq \hat{Q}\leq Q$, so the last cycle is executed at most two times). These computations require $2M(n)$ time (ignoring shifts and subtractions and the precalculus $\left\lfloor\frac{2^{2n}}{N}\right\rfloor$). Therefore the Barrett algorithm reduces $MR(2n,n)$ to $2M(n)$. A further observation is that in the multiplication $\left\lfloor\frac{x}{2^n}\right\rfloor \left\lfloor\frac{2^{2n}}{N}\right\rfloor$ only $\sim n$ upper bits have to be computed while in the multiplication $\hat{Q}N$ only $\sim n$ lower bits have to be computed. This observation however, is useful if multiplications are done using schoolbook algorithm (in this case both moltiplications time can be halved) while is practically useless if the multiplications are done with Karatsuba or Toom Cook\footnote{The possibility to use subquadratic algorithm for multiplications is one of the main attractive features of Barrett algorithm. This cannot be done in Montgomery algorithm where a sequence of additions is instead performed. For not very large values of $n$ (i.e. $n=1024$) it is still convenient perform multiplications via the schoolbook method and Montgomery algorithm turns to be about $20\%$ faster than Barrett one.}. If we use these algorithms we can however speed up modular reduction in another way:
\begin{itemize}
\item we can use some additional precalculus in order to reduce a modular reduction from $2n$ to $n$ bits to another (shorter) one from $m$ to $n$ bits ($n<m<2n$);
\item then the generic modular reduction $x\ mod(N)$ where the bitlength of $x$ is $m$ and the bitlength of $N$ is $n$ can be performed by a "modified" Barrett algorithm:
\end{itemize}

\begin{array}{l}
\hat{Q}\leftarrow \left\lfloor  \frac{\left\lfloor \frac{x}{2^n} \right\rfloor \left\lfloor\frac{2^{m+n}}{N}\right\rfloor}{2^m} \right\rfloor\\
x\leftarrow x-\hat{Q}N\\
while\ x\geq  N\\
\{ x\leftarrow x-N\}
\end{array}

This algorithm requires $M(m)+M(m,n)$ time.

We now, following such observations,  give the detailed description of two different methods to efficiently perform a modular reduction from $2n$ to $n$ bits.

\subsection[First method to compute the modular reduction]{First method to compute the modular reduction}\label{me1}

The idea is to divide the number $x$, that has bit-length $2n$ bits,  as follows:
$$x=x_12^{\frac{3}{2}n}+x_22^n+x_3$$
with $x_1,x_2 \sim \frac{n}{2}\ bits$ and $x_3 \sim n\ bits$.
Then:
$$xmod(N)\equiv \left(x_1\left(2^{\frac{3}{2}n}mod(N)\right)+x_22^n+x_3\right)mod(N)$$

It is possible to pre-compute $2^{\frac{3}{2}n}mod(N)$. 

We observe that $x_12^{\frac{3}{2}n}mod(N)$ costs $M(\frac{n}{2},n)\simeq 2M(\frac{n}{2})$ (neglecting the pre-computation time). Then the cost of $MR(\frac{3}{2}n,n)$ is $M(\frac{n}{2})+M(\frac{n}{2},n)\simeq 3M(\frac{n}{2})$.

Therefore the total cost of $MR(2n,n)$ is $5M(\frac{n}{2})$.

This method is more efficient than Barrett algorithm which requires $2M({n})\simeq 6M(\frac{n}{2})$, see (\ref{eq1}), with a maximum theoretical gain of $\frac{1}{6}\simeq 16,7\%$.

\subsection{Method to compute a scalar product}\label{oss1}

If we have to compute a scalar product with big integers, $(a,b)$ and $(c,d)$:
$$(a,b)\cdot (c,d)=ac +bd$$
we have to compute two multiplications. Assuming to know $ab$ and $cd$, we observe that:
$$(a-d)\cdot(b-c)=ab-ac-bd+cd$$
$$\Rightarrow ac+bd=(d-a)(c-b)+ab+cd$$
so that the scalar product can be computed with only one multiplication (we assume that the computational cost of addition and subtraction is negligible respect the cost of multiplication).

\subsection[Second method to compute the modular reduction]{Second method to compute the modular reduction}\label{met2}

In this section we illustrate a method that reduces  $MR(2n,n)$ to $8M(\frac{n}{3})$. Let $x$ be an integer of bit-length $2n$, then:

$$x=x_12^{\frac{5}{3}n}+x_22^{\frac{4}{3}n}+x_3$$
with $x_1,x_2 \sim \frac{n}{3}$ bits and $x_3\sim \frac{4}{3}n$ bits. Therefore: 

$$xmod(N)=$$
$$=\left(x_1\left(2^{\frac{5}{3}n}mod(N)\right)+x_2\left(2^{\frac{4}{3}n}mod(N)\right)+x_3\right)mod(N)$$
$$=(x_1R_1+x_2R_2+x_3)mod(N)=Cmod(N)$$
where $R_1$ and $R_2$ have bitlength $n$ and can be pre-computed. If we set:
$$R_1=R_1^H2^{\frac{2}{3}n}+R_1^I2^{\frac{n}{3}}+R_1^L$$
$$R_2=R_2^H2^{\frac{2}{3}n}+R_2^I2^{\frac{n}{3}}+R_2^L$$
the amount $x_1R_1+x_2R_2$ can be computed via three scalar products:
$$(x_1-R_1^M)(R_2^M-x_2)=x_1R_1^M+x_2R_2^M-x_1x_2-R_1^MR_2^M$$
where $M=H,I,L$. Since $R_1^MR_2^M$ can be pre-computed then it possible to compute $C$ in $4M(\frac{n}{3})$. Now we use Barrett algorithm to compute last modular reduction which requires $1M(\frac{n}{3},n)\simeq 3M(\frac{n}{3})$ and $1M(\frac{n}{3})$. Then the total cost is $8M(\frac{n}{3})$ that is less than $2M(n)\simeq 10M\left (\frac{n}{3}\right)$ by (\ref{eq1}), with a maximum theoretical gain of $20\%$.

\subsection{Implementation of the basic iteration of QUAD}

To compute the value of a quadratic polynomial in $n$ indeterminates in GF(2) we have:\footnote{There are no terms with $x_i^2$ since $x_i^2=x_i$ in $\mathbb{Z}_2$}
$$P({\bf x})=\sum_{1\leq i< j\leq n}{a_{i,j}x_ix_j}+\sum_{1\leq i\leq n}{b_ix_i}+c$$
We observe that the addition and  xor give the same result in $\mathbb{Z}_2$. The computation of the iteration $f$ consists in the computation of $2n$ (we fixed $k=2$) quadratic polynomial values at point ${\bf x}$ and can be efficiently realized in $\mathbb{Z}_2$ by xoring the columns of the coefficients of $2n$ polynomials $P_1({\bf x}),...,P_{2n}({\bf x})$; more precisely we have to xor the columns of coefficients $b_i^h$, $1\leq h\leq 2n$,  for which $x_i=1$, $a_{i,j}^h$ for which $x_ix_j=1$. Such columns can be arranged in words of $d$ bits ($d=32$ or 64 bits depending on the o.s.).
The average number of xor for each iteration for a MQ system and each bit outputted is:
$$\frac{n+\frac{1}{2}\bigg(\begin{matrix} n \\ 2 \end{matrix} \bigg)}{d}$$

A faster implementation of QUAD is possible if one makes some precomputations. To give an idea of how this is possible imagine that $n=2lm$, that is the $n$ indeterminates can be arranged in $m$ blocks of $2l$ indeterminates each. As the index $i$ varies in the same block of $2l$ indeterminates all values $x_i$ and $x_ix_j$ ($i,j$ belonging to the same block) are determined, therefore one can precompute the xor of the columns according to these values and store the new columns in a new matrix ($2^{2l}-1$ columns have to be stored for each block). Then we have to consider values $x_i$ and $x_j$ where $i,j$ belong to two different blocks. Then one can split each one of the two blocks in two parts and for $i$ and $j$ spanning such half blocks one can xor the corresponding coefficient columns according to the value $x_ix_j$. We have to store $4(2^l-1)^2$ new columns for each couple of different blocks.

Then the implementation of QUAD now reduces to xor one column for each block and 4 columns for each couple of different blocks. The average number of xor for each bit outputted will now be:
$$\frac{2\left( m+4\left(\begin{matrix} n \\ 2 \end{matrix} \right)\right)}{d}$$
For a more detailed description of the procedure see[]. 

In the numerical implementation we will use $n=160$ (the suggested value by  Berbain,  Gilbert and Patarin) and beside the classical implementation also the implementation we just described, taking $l=2$ or $l=4$. In order to highlight the trade off between the memory needed to store the coefficients of the system and the number of xor/bit required we just have this table (assuming the word length $d=64$):

\begin{table}[h]
\begin{center}
\begin{tabular}{|c|c|c|}
\hline
\hline {{\bf QUAD implementation}} & {\bf Memory} & {\bf xor/bit} \\
\hline { Classical  }& 0,4 Mbit &    102 \\
\hline { $l=2$} & 10 Mbit  &  56 \\
\hline  { $l=4$}& 54 Mbit &  18 \\
\hline
\hline
\end{tabular}
\caption[]{}
\label{tabella10}
\end{center}
\end{table}

\subsection{A faster variant of RSAPRG and QUAD}\label{pre}

In \cite{Articolo1} we also described a way to increase the throughput of a TCSPRBG by replacing the output $y_n$ with $w(y_n)$ where $w$ is a suitable one way function that both expands the number of bits outputted and is fast to compute.

We tested the speed of calculation of a concrete proposal for such function and estimated the throughput that can be achieved by RSAPRG and QUAD modified in this way.

\section{Numerical implementations}

\begin{figure}
\centering
\includegraphics[width=3.4in]{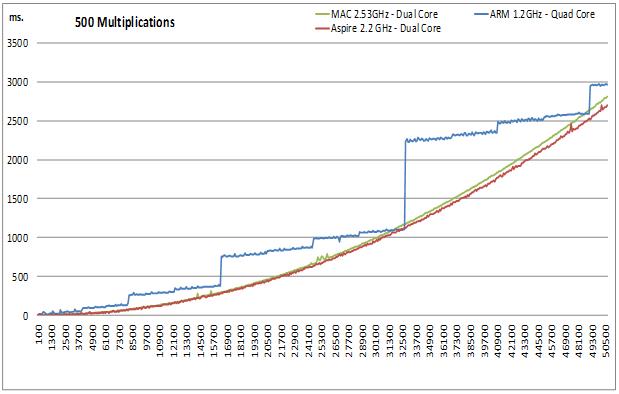}
\caption{Comparison of the performance of multiplication of big integer for an Aspire 2.2GHz, a Mac 2.53GHz and a Google Nexus 1.2 GHz}\label{figura1}
\end{figure}

We have exploited the fact that the BigInteger class of Java language is capable of handling integers that have a sufficient bitlength to be used in cryptographic applications. We initially wrote a simple benchmark program that measures the execution time of $M(n)$. Our program was compiled and run both on devices with Android o.s. and on devices with Windows or Mac o.s., namely:
\begin{itemize}
\item Mac Pro - Intel core 2 duo  P8700 - 2,53 GHz;
\item Acer Aspire - Intel Pentium B960 - 2,2 GHz;
\item Google Nexus 7 - Nvidia Tegra 3 - 1,2 GHz\footnote{our code, anyway, does not make any use of parallel computations.}.
\end{itemize}
The plot obtained for these machines is reported in figure \ref{figura1}.
When our program was run on devices with Windows o.s. or Mac o.s., we can observe that the graph of $M(n)$ is practically quadratic in $n$ (therefore the BigInteger class seems to use neither Karatsuba algorithm nor the Tom Cook algorithm). On the Google Nexus 7 the execution times present a much more singular trend. There are significant jumps (the execution time doubles) corresponding to bitlengths given by powers of 2. There are also reduced jumps corresponding to $3\cdot 2^k$ with $k$ integer. The execution time triples between $n$ and $2n$, therefore the program seems to use Karatsuba algorithm or Tom Cook algorithm. To support this thesis we found in internet some modified versions of BigInteger class that implement Karatsuba algorithm and Toom Cook algorithm for multiplication. One of these is available at the website "futureboy.us/temp/BigInteger.java" and it has been written by Alan Eliasen .

\subsection{Numerical results for RSAPRG}

Then we fully developed Java code to implement the basic iteration for RSAPRG. 
For benchmark programs of RSAPRG we fixed $n=6144$.
Since the BigInteger class of Java has a different behavior on Windows and Android devices, we will consider them separately. 

\subsubsection{RSAPRG - Windows or Mac devices}

In this case BigInteger class implements the multiplication in a classical way then  it can be applied one or two level of Karatsuba algorithm to speed up multiplications time. In practice it is convenients to apply one level of Karatsuba algorithm (see table \ref{tabella1})\footnote{Time is averaged on 10 runs with random chosen data. The same is true until up to table \ref{tabella7}.}.

\begin{table}[h]
\begin{center}
\begin{tabular}{|c|c|c|}
\hline
\hline {\bf $10^3$ Multiplications} & {\bf Mac Pro} & {\bf Aspire} \\
\hline Classical & 85 &  80   \\
\hline Karatsuba 1 level & 73 &  67 \\
\hline Karatsuba 2 levels& 73 &  66 \\
\hline Toom Cook 3& 77 &  66 \\
\hline
\hline
\end{tabular}
\caption[]{Time in ms to compute $10^3$ multiplications (each multiplicand has bitlength $n=6144$). }
\label{tabella1}
\end{center}
\end{table}

From table \ref{tabella2} we can see that $MR(2n,n)$ can be more than twice as fast using method 2 with respect to the calculation of remainder by BigInteger class. Part of these gain is due to the fact that BigInteger class does not implement subquadratic algorithms: in this case assumption (\ref{eq1}) modifies in this way:
$$M(n)\simeq \left ( \frac{n}{2}\right )\simeq 9M\left ( \frac{n}{3} \right )$$
Therefore the maximal theoretical gain for method 1 is $5M\left ( \frac{n}{2}\right )$ for method 1 compared to $8M\left ( \frac{n}{2}\right )$ for Barrett, $8M\left( \frac{n}{3}\right)$ compared to $18M\left ( \frac{n}{3}\right )$ for Barrett. This is consistent with results from table \ref{tabella2}.

\begin{table}[h]
\begin{center}
\begin{tabular}{|c|c|c|}
\hline
\hline {\bf $10^3$ Modular reduction} & {\bf Mac Pro} & {\bf Aspire} \\
\hline Classical & 186 &  182   \\
\hline Barrett (with 1 level Karatsuba) & 156 &  150 \\
\hline First Method & 114 &  107 \\
\hline Second Method & 94 &  84 \\
\hline
\hline
\end{tabular}
\caption[]{Time in ms to compute $10^3$ modular reductions.}
\label{tabella2}
\end{center}
\end{table}

In table \ref{tabella3} we report the execution time for the basic iteration of RSAPRG, like described in section 3.1. Applying one level of Karatsuba algorithm for multiplications and implementing modular reductions following methods described in sections 3.2 and 3.4 yields a time saving up to more than $40\%$.

\begin{table}[h]
\begin{center}
\begin{tabular}{|c|c|c|}
\hline
\hline {\bf $10^3$ ${\bf x^9mod(N)}$} & {\bf Mac Pro} & {\bf Aspire} \\
\hline Classical & 955 &  953   \\
\hline First Method & 688 &  629 \\
\hline Second Method & 602 &  533 \\
\hline
\hline
\end{tabular}
\caption[]{Time in ms to compute $10^3$ exponentiations ($e=9$).}
\label{tabella3}
\end{center}
\end{table}

Taking in to account that at each iteration 2196 bit can be outputted, results from table \ref{tabella3} correspond to a throughput for RSAPRG of 4,1 Mbit/sec for Aspire and 3,6 Mbit/sec for Mac Book Pro.

\subsubsection{RSAPRG - Android devices}

The BigInteger class of Java on Android devices implements the multiplication using Karatsuba and Toom Cook 3 algorithm. Therefore we did not implement any level of Karatsuba algorithm. 

When we apply our methods to speed up modular reduction (see \ref{tabella4}), we  observe that results are far from those expected, in fact times don't fall below the classical times.

We think that the reason is that a lot of time is lost to access memory when we split the operands of same multiplications in several pieces. For example the time to compute additions depends very little on the bitlength of operands (see table \ref{tabella5}). The picture changes if we consider modular reductions $MR(2n,n)$ for larger $n$. When $n=24000$ or $48000$ our methods are effective also with respect to Barrett algorithm (see table \ref{tabella6}). 

\begin{remark}
We also optimized code to always multiplicate operands of the same bitlength, this anyway makes things even worse for our methods when $n=6144$.
\end{remark}

\begin{remark}
The gain of second method with respect to Barrett algorithm in table \ref{tabella6} for $n=49152$ is greater than $20\%$, that is the gain forecast based on section 3.4. This should be due to the jump effect highlighted in remarks to figure \ref{figura1}.
\end{remark}

\begin{table}[h]
\begin{center}
\begin{tabular}{|c|c|}
\hline
\hline {\bf $10^3$ Modular reductions} & {\bf Google Nexus 7}  \\
\hline Classical & 546   \\
\hline Barrett    & 579 \\
\hline First Method  & 569  \\
\hline Second Method  & 860 \\
\hline
\hline
\end{tabular}
\caption[]{Time in ms to compute $10^3$  $MR(2n,n)$ ($n=6144$) on Google Nexus 7}
\label{tabella4}
\end{center}
\end{table}

\begin{table}[h]
\begin{center}
\begin{tabular}{|c|c|}
\hline
\hline {\bf Google Nexus 7} & {\bf 50000 Additions}  \\
\hline $n=200$ & 688   \\
\hline 2000 & 633 \\
\hline 4000 & 679  \\
\hline 6000 & 770 \\
\hline 8000 & 743 \\
\hline 10000 & 763 \\
\hline
\hline
\end{tabular}
\caption[]{Time in ms to compute 50000 additions with operand of different bitlength  on Google Nexus 7}
\label{tabella5}
\end{center}
\end{table}

\begin{table}[h]
\begin{center}
\begin{tabular}{|c|c|c|c|c|}
\hline
\hline {\bf  Mod. reductions} & {\bf n=6144} & {\bf 12288} & {\bf 24576} & {\bf 49152}  \\
\hline Classical & 520 & 1755 & 6528 & 25342   \\
\hline Barrett  & 558 & 1362 & 3673 & 10625 \\
\hline First Method & 709 & 1396 & 3462 & 9444  \\
\hline Second Method & 943 & 1477 & 2935 & 7274 \\
\hline
\hline
\end{tabular}
\caption[]{Time in ms to compute $10^3$ modular reductions with operand of different bitlength  on Google Nexus 7}
\label{tabella6}
\end{center}
\end{table}

\begin{table}[h]
\begin{center}
\begin{tabular}{|c|c|c|c|c|}
\hline
\hline {\bf  ${\bf x^9mod(N)}$} & {\bf n=6000} &{\bf 12288} & {\bf 24576} & {\bf 49152}\\
\hline Classical & 2998 & 9945 & 34029 & 131817   \\
\hline Barrett  & 3126 & 8072 & 22779 & 72533 \\
\hline First Method & 3965 & 8421 & 21840 & 68012  \\
\hline Second Method & 4975 & 8629 & 19946 & 58948 \\
\hline
\hline
\end{tabular}
\caption[]{Time in ms to compute $10^3$ exponentiations with operands of different bitlength on Google Nexus 7}
\label{tabella7}
\end{center}
\end{table}

The time nedeed for exponentiation is reported in table \ref{tabella7} and corresponds to a throughput of 0,7 Mbit/sec for $n=6144$ (using classical algorithm).

\begin{table}[h]
\begin{center}
\begin{tabular}{|c|c|c|c|}
\hline
\hline {\bf QUAD} & {\bf Mac Pro} & {\bf Aspire} & {\bf Google Nexus7}\\
\hline {\bf Classical}& 445 &  223 & 4762  \\
\hline {\bf $l=2$}& 308 &  296 & 3990  \\
\hline {\bf $l=4$}& 447  &  295 & 2673  \\
\hline
\hline
\end{tabular}
\caption[]{Time in ms to compute $10^4$ basic iterations of QUAD ($n=160$)}
\label{tabella9}
\end{center}
\end{table}

\begin{table}[h]
\begin{center}
\begin{tabular}{|c|c|c|c|}
\hline
\hline {\bf Function $w$} & {\bf Mac Pro} & {\bf Aspire} & {\bf Google Nexus7}\\
\hline {$10^4$ iterations} & 302 &    265 & 4098  \\
\hline
\hline
\end{tabular}
\caption[]{Time in ms to compute $10^4$ evaluations of function $w$.}
\label{tabella11}
\end{center}
\end{table}

\subsection{Numerical results for QUAD}

In table \ref{tabella9} we report the time needed to perform $10^4$ basic iterations of QUAD taking $n=160$, as suggested by Berbain, Gilbert and Patarin. The rows with $l=2$ and $l=4$ correspond to the attempt to speed up the implementation via some precomputation as described in section \ref{pre}. We can deduce that for Google 7 we have some concrete gain, for Mac we have a gain only with $l=2$ while for Aspire we have no gain at all.
The best results of table \ref{tabella9} correspond to a throughput of 7,1 Mbit/sec for Aspire, 5.2 Mbit/sec for Mac and between 0,5 and 0,6 Mbit/sec for Google Nexus 7.

\subsection{Numerical results for function ${\bf w}$}

In the table \ref{tabella11} we report the time needed to perform $10^4$ iterations of a concrete example of function $w$ we considered in \cite{Articolo1}. Such function expands 768 bits to an output of 4096 bits.

\begin{table}[h]
\begin{center}
\begin{tabular}{|c|c|c|c|c|} \hline
\multicolumn{1}{|c|}{} &
\multicolumn{2}{c|}{{\bf RSAPRG}} &
\multicolumn{2}{c|}{\bf QUAD}\\ \hline
  & {\bf 1 time} & {\bf 2 times} & {\bf 1 time} & {\bf 2 times} \\ \hline
 {\bf Aspire} & 19,2 & 61,6 & 30,6 & 79,4\\ \hline
 {\bf Mac} & 17,0 & 54,3 & 23,0 & 64,42\\ \hline
 {\bf Nexus 7} & 2,8 & 6,0 & 2,4 & 5,6\\ \hline
\end{tabular}
\caption[]{Calculated throughput of modified versions of RSAPRG and QUAD in ms.}
\label{tabella12}
\end{center}
\end{table}

In \cite{Articolo1} we suggest to apply function $w$ one or two times to the output of a CSPRBG in order to obtain the final output. In table \ref{tabella12} we report the calculated throughput of such modified versions of RSAPRG and QUAD.

\section{Conclusion}
We showed that it is possible to develop implementations in Java for two well known "provable secure" CSPRBG, namely RSAPRG and QUAD. The throughput rate achieved for RSAPRG seems to be better than the one estimated by authors in \cite{RSAPRG}, section 5, table 1 (0,67 Mbit/sec on a Pentium 4). The throughput rate achieved for QUAD is similar to the values decleared in \cite{1}. The maximum speed we achieved is 2453 cycles/byte on Acer Aspire compared to 2081 cycles/byte achieved in \cite{2} on AMD opteron 2,2 GHz, L2 cache 1024 kb. While it is true that we tested our programs on more recent processors, Java implementations run on a virtual machine. We also developed several techniques to speed up such implementations, pointing out some differences in the behaviour of BigInteger class and in handling of memory access under different o.s. We eventually applied a modification of such CSPRBGs we proposed in \cite{Articolo1}, showing we can effectively increase the throughput rate by one order of magnitude. In \cite{Articolo1} we extensively discuss security implications of such modification.

\begin{IEEEbiography}{Antonio Corbo Esposito}
Biography text here.
\end{IEEEbiography}

\begin{IEEEbiographynophoto}{Antonio Corbo Esposito}
Biography text here.
\end{IEEEbiographynophoto}

\begin{IEEEbiographynophoto}{ Antonio Corbo Esposito}
Biography text here.Biography text here.
\end{IEEEbiographynophoto}

\end{document}